\documentclass[12pt]{iopart}

\usepackage{graphicx}

\usepackage{makecell}

\newcommand*{\OrigAA}{}
\let\OrigAA\AA
 
\renewcommand*{\AA}{%
  {\fontfamily{ptm}%
  \selectfont%
  \OrigAA%
  \selectfont}%
}

\begin{document}

\title[21 cm line profile of hydrogen atom in ISM]{Analysis of the Absorption Line Profile at 21 cm for the Hydrogen Atom in Interstellar Medium}

\author{D. Solovyev}

\address{Department of Physics, St. Petersburg State University 198504, St. Petersburg, Russia}

\eads{\mailto{solovyev.d@gmail.com}}

\begin{abstract}
The paper analyzes the absorption line profile at 21 cm for the hydrogen atom in the interstellar medium. The hydrogen atom is treated as a three-level system illuminated by a powerful light source at neighboring resonances corresponding to the hyperfine splitting of the ground state and Ly$_\alpha$ transition. The field acting upon the resonances gives rise to physical processes, which can be explained as interfering pathways between different transitions. The paper considers particular cases when the 21 cm line profile is substantially modified by the Ly$_\alpha$ transition. A correction to the optical depth is introduced as a result of theory. It is shown that the correction can be considerable and should be taken into account when determining the column density of hydrogen atoms in the interstellar medium. The paper also deals with the effects of none-Doppler broadening and frequency shift.
\end{abstract}
\maketitle
%\ioptwocol

\section{Introduction}

Investigation of the interstellar medium (ISM) is of an extreme importance for understanding the factors and mechanisms responsible for the formation of gas clouds and dust complexes and their part in the evolution of stars. Measurements dealing with radio-loud sources can provide information on the structure and physical conditions of galaxies. The observations of hydrogen clouds are the most probable candidate as they cover greater part of the interstellar gas. The $21$ cm absorption line in hydrogen atom has a special meaning in investigations of the type, allowing the determination of the distribution and kinematical properties of the neutral hydrogen. Since the direct imaging is restricted by large observing time requirements at all wavebands, the optical imaging studies are complicated \cite{Kan}. Inasmuch as the interstellar medium is a transparent for radio frequencies, this makes the HI 21 cm line the obvious target for the studies.

At the same time, the damped Lyman-$\alpha$ (DLA) systems are of particular interest as the high atomic hydrogen HI column density absorbers. The Lyman $\alpha$ absorption can be used as a very sensitive probe of the HI column in clusters of galaxies \cite{Laor}. The large cross-section for the Ly$_{\alpha}$ transition makes the technique the most sensitive method for detecting baryons at any redshift \cite{Rauch}. The Lyman-$\alpha$ (whose profile itself is controlled by the kinetic temperature) provides a more effective coupling between the spin temperature and the kinetic temperature for the high density cloud illuminated by a powerful source of light \cite{Rees}. The damping wings of the Lorentzian component of the absorption profile become possible to be detected from about the $\sim 10^{19}$ $cm^{-2}$ column density, attaining their maximum in the "damped Ly$_\alpha$ systems" \cite{Rauch}. At low densities of the ISM, collisions are inefficient for lowering the spin temperature. If Lyman alpha radiation penetrates the HI without heating it, it can actually lower the spin temperature so that the 21 cm line becomes a stronger absorption feature. Thus, the $21$ cm absorption and Ly$_{\alpha}$ absorption line profiles provide two independent tools for the investigation of ISM. Yet, investigation in the absorption $21$ cm and Ly$_{\alpha}$ lines cannot be considered separately. For example, author of \cite{Varsh} has considered the populations of the magnetic sublevels for the hyperfine splitting of the ground state $2s_{1/2}$ in hydrogen atom, whereas \cite{Varsh} investigated the atomic orientation depending on the intensity, spectrum, angular distribution, and polarization of the incident optical and radio emission.

In view of the above, the paper focuses on the interstellar neutral hydrogen atom subjected to an external field at frequencies of hyperfine energy splitting of the ground state, $21$ cm, and the Ly$_{\alpha}$ transition. Description of such atom-field system can be reduced to the examination of the three-level atom. The main constraints in our analysis arise due to describing the hyperfine structure of the ground state without the account for the fine and hyperfine structure of the excited $2p$ state. However, the approximation can be justified by a large value of the Ly$_{\alpha}$ transition rate and approximate equivalence of the one-photon $1s\leftrightarrow 2p$ transition probabilities between the fine and hyperfine sublevels in the hydrogen atom.% Moreover, observation of the absorption line on the $21$ cm transition makes it unnecessary to consider the degenerate excited levels in detail. We should also note that the accuracy restriction imposed on the atom-field system for the interstellar medium stems from the accuracy of measuring the density flux at the corresponding frequency.

\begin{figure}[hbtp]
	\centering
	\includegraphics[scale=0.18]{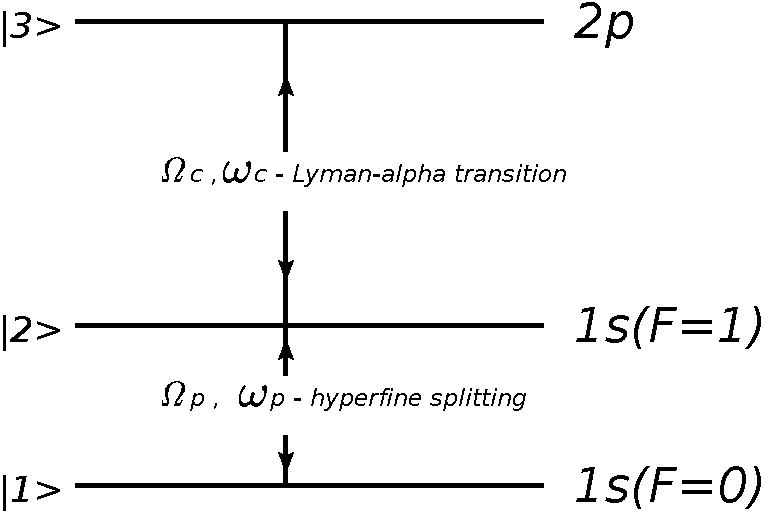}
	\caption{Schematic representation of energy levels in a hydrogen atom. The lower states correspond to the ground state with the hyperfine sublevels for the total angular momenta $F=1$ and $F=0$. The frequency of the transition between the hyperfine sublevels is  $21$ cm. The upper state is the $2p$ excited state corresponding to the Lyman-$\alpha$ transition. $\Omega_p$ and $\Omega_c$ denote the corresponding Rabi frequencies for the probe and controlled fields of the external source.}
	\label{Fig1}
\end{figure}

The paper provides a detailed analysis of the absorption profile at 21 cm line for the hydrogen atom in the interstellar medium. To this end,  the atom-field system is described within the framework of density matrix formalism \cite{Boyd}. The hydrogen atom is treated as a three-level ladder (cascade) system: two hyperfine sublevels of the ground atomic state and excited $2p$ state, see Fig.~{\ref{Fig1}}.
The unperturbed 21 cm line profile is shown to arise at a negligible field strength of the neighboring transition ($\Omega_c\rightarrow 0$). However, the profile can be modified significantly by the Ly$_\alpha$ transition induced in the field of a powerful light source. In this case, the absorption process cannot be thought of as a one-photon transition between the hyperfine sublevels of the ground state in the hydrogen atom. The developed theory allows finding a correction to the optical depth. The correction correlates directly with the determination of the coulmn density of hydrogen atom in the interstellar medium. The paper presents certain events where the correction is of significance, some of them just illustrating the importance of the analysis.

%It is shown that the correction to the optical depth arising via the adjacent Ly$_\alpha$ resonance can as high as the experimental accuracy $\approx 0.1\%$ of the hydrogen column density, see, for example, \cite{Fox}. Basing on the formalism, we consider the three-level  scheme of hydrogen atom. We show that the Lorentz line profile, which corresponds to the one-photon absorption process, can be obtained only in case of the negligible field strength on the neighboring transition. In most frequent cases, the absorption line profile is formed via additional physical processes occurring within the system in the presence of the field, but not the "one-photon absorption", which leads to overestimation of the column density of hydrogen atoms. Finally, the line broadening and frequency shift arising for such atom-field system is discussed throughout the paper.

\section{Correction to the optical depth via the evaluation of three-level $\Xi$ scheme}
\label{opt}
%To describe the ladder scheme of the hydrogen atom, we employ the density matrix formalism. The details of the theory can be found in a number of textbooks, see, for example, \cite{Boyd}. Recently, this formalism was applied to the descritpion of the cosmological recombination era of universe \cite{SDP}. It was found that a correction on the order of $1\%$ can be derived for the optical depth.

Recently, the density matrix formalism was applied to study the effect of Electromagnetically-Induced Transparency (EIT) in hydrogen atom under the conditions corresponding to the recombination era of the universe \cite{SDP,SD}. It was shown that the EIT phenomenon could lead to corrections on the order of $1\%$ in the observed cosmic microwave background (CMB). To study the absorption line profile corresponding to the $21$ cm transition in hydrogen atom, the present work draws on the same formalism. 

Theoretical description of an atom within the three-level cascade $\Xi$-scheme approximation can be found in \cite{Whitley,Gea-Ban}, which employed the density matrix formalism. The example of the rubidium atom was used to investigate the EIT phenomenon in \cite{Wiel}, where the physical picture of the effect was rendered in terms of interfering multiphoton transition pathways. Basing on the formalism given in \cite{Whitley}-\cite{Wiel}, the influence of EIT phenomenon on the optical depth determination for the interstellar hydrogen atom is investigated. The three-level $\Xi$ scheme in hydrogen atom with the energy levels: state $|1\rangle$ is the lower and state $|2\rangle$ is the upper hyperfine sublevels of the ground $1s$ state, with state $|3\rangle$ representing the excited $2p$ level in the hydrogen atom. The absorber is assumed to be subjected to the "probe" and "controlled" fields with the corresponding frequencies $\omega_p\rightarrow\omega_{21}=|1\rangle\leftrightarrow|2\rangle$ and the Ly$_{\alpha}$ transition $\omega_c\rightarrow\omega_{32}=|2\rangle\leftrightarrow|3\rangle$.

The set of equations for the ladder scheme is
\begin{eqnarray}
\label{1}
\rho_{21}=\frac{i/2\left(\Omega_p(\rho_{22}-\rho_{11})-\Omega_c^*\rho_{31}\right)}{\gamma_{21}-i\delta_p}\, ,
\\
\nonumber
\rho_{32}=\frac{i/2\left(\Omega_c(\rho_{33}-\rho_{22})+\Omega_p^*\rho_{31}\right)}{\gamma_{32}-i\delta_c}\, ,
\\
\nonumber
\rho_{31}=\frac{i/2\left(\Omega_p\rho_{32}-\Omega_c\rho_{21}\right)}{\gamma_{31}-i(\delta_p+\delta_c)}\, ,
\\
\nonumber
\rho_{22}=\frac{i}{2\Gamma_2}(\Omega_p^*\rho_{21}-\Omega_p\rho_{12})\, ,
\\
\nonumber
\rho_{33}=\frac{i}{2\Gamma_3}(\Omega_c^*\rho_{32}-\Omega_c\rho_{23})\, ,
\end{eqnarray}
where detunings for the probe and controlled fields are $\delta_p=\omega_p-\omega_{21}$, $\delta_c=\omega_c-\omega_{32}$, respectively, and $\omega_{21}$, $\omega_{32}$ represent the exact values of corresponding transitions. The Rabi frequencies are denoted with $\Omega_c=2d_{32}E_c/\hbar$ and $\Omega_p=2\mu_{21}B_p/\hbar$. Since the transition between the hyperfine sublevels corresponds to the magnetic dipole M1 emission/absorption, Rabi frequency $\Omega_p$ is written in terms of magnetic field strength $B_p$ and magnetic moment $\mu_{ij}$. $E_c$ represents the electric field strength for Ly$_{\alpha}$ transition and $d_{ij}$ is the dipole matrix element. The wave functions for $|3\rangle$, $|2\rangle$, and $|1\rangle$ states can be taken as the solution of Schr\"odinger equation. In the absence of collisions, $\gamma_{ij}=(\Gamma_i+\Gamma_j)/2$, where $\Gamma_i$ is the natural width of the $i$th level. The set of equations (\ref{1}) is written in the steady-state and rotating wave approximations, see \cite{Whitley}-\cite{Wiel}.

With the use of Eqs. (\ref{1}), the monochromatic absorption coefficient at frequency $\omega_{ij}$ can be defined as
\begin{eqnarray}
\label{2}
k=\frac{N d_{ij}^2\omega_{ij}}{2\varepsilon_0\Omega_{ij}}Im\big\{\rho_{ij}\big\}\, ,
\end{eqnarray}
where $\varepsilon_0$ is the vacuum permittivity, $N$ is the number of atoms and $\Omega_{ij}$ is the corresponding Rabi frequency. Taking into account the relation $k=\tilde{k}\phi(\omega)$, where $\tilde{k}$ is the integrated line absorption coefficient and $\phi(\omega)$ is the normalized line profile, the monochromatic optical depth is
\begin{eqnarray}
\label{3}
d\tau(\omega_{ij})=-\tilde{k}\phi(\omega_{ij})dl=-\tau\phi(\omega_{ij})\frac{dl}{L}\, ,
\end{eqnarray}
where $l$ is the distance along the ray \cite{Sobolev}. %from the emission point and $L$ is the Sobolev length $L=\sqrt{3k_B T_M/m_{atom}}/| \it{v}'|$, see \cite{Sobolev}, $k_B$ is the Boltzmann constant, $T_M$ is the gas temperature, $m_{atom}$ is the atomic mass and $\it{v}'$ is the velocity gradient. Optical depth characterizes the medium opacity for the radiation passing through it, i.e. the intensity at frequency $\omega$ can be expressed as $I_\omega = I_\omega(0)exp(-\tau(\omega))$. Optical depth is related to the column density of atoms, $N$, via $\tau(\omega) = \sigma(\omega) N$, where $\sigma(\omega)$ is the absorption cross-section in line $\omega$. Thus, corrections on the order of $0.1\%$ (experimental accuracy of the hydrogen column density determination \cite{Fox}) to the optical depth can be essential for the determination of  column density $N$. %Thus, the imaginary part of the density matrix element $\rho_{21}$ is of the particular interest.

In ordinary case the monochromatic optical depth corresponds to the one-photon resonant process that reduces to the evaluation of the two-level atomic system. Then the one-photon absorption process is described by the Lorentz line profile:
\begin{eqnarray}
\label{3a}
Im\{\rho_{21}^{(0)}\}=-\frac{\gamma_{21}\Omega_p/2}{\delta_p^2+\gamma_{21}^2}\, ,
\end{eqnarray}
where $\delta_p$ can be considered as a variable. %The accounting for the Doppler effect leads to the Voigt profile, which is given by the convolution of Lorentz and Gauss profiles. It is the Voigt profile that is used for fitting of the absorbtion line observed in astrophysical investigations of interstellar medium. However, below, the modification of natural Lorentzian line profile (Lorentzian) due to the presence of external field on adjacent resonance is investigated. 
A more accurate solution of Eqs. (\ref{1}) corresponds to accounting of the second field acting upon the adjacent resonance. Then, in the limit of the weak "probe" field \cite{Gea-Ban}, matrix element $\rho_{21}$ in the first order of the "probe" field and in all orders of the "control" field is
\begin{eqnarray}
\label{1a}
\rho_{21}=\frac{i\Omega_p/2}{i\delta_p-\gamma_{21}+\frac{\Omega_c^2/4}{i(\delta_p+\delta_c)-\gamma_{31}}}.
\end{eqnarray}
%The deduction of the Lorentz multiplier Eq. (\ref{3a}) (for the negligible values $\Omega_c$, for example) in (\ref{1a}) allows defining the integrated absorption coefficient. Thus, the monochromatic optical depth Eq. (\ref{3}) can be defined at the corresponding frequency. 
Expression (\ref{1a}) depends on field parameters $\Omega_c$ and $\delta_c$ and reduces to Eq. (\ref{3a}) in the limit $\Omega_c\rightarrow 0$, i.e. when the influence of the field on adjacent resonance is negligible. In this case, the corrections to 'ordinary' determination (\ref{3a}) can be found via the series expansion in $\Omega_{p(c)}$ at zero detunings $\delta_{p(c)}$. Then the transition amplitudes associated with $|1\rangle\rightarrow |2\rangle$ and $|2\rangle\rightarrow |3\rangle$ pathways result in the destructive interference and the reduction of the total probability that a probe photon will be absorbed \cite{Wiel}. 

\begin{figure}[hbtp]
	\centering
	\includegraphics[scale=0.5]{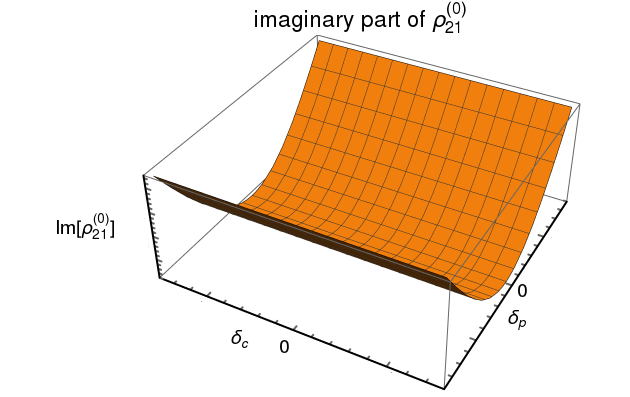}
	\caption{Schematic image of the line profile corresponding to $Im\{\rho_{21}^{(0)}\}$, Eq. (\ref{3a}), for different values of detunings $\delta_p$ and $\delta_c$.}
	\label{Fig2}
\end{figure}

\begin{figure}[hbtp]
	\centering
	\includegraphics[scale=0.5]{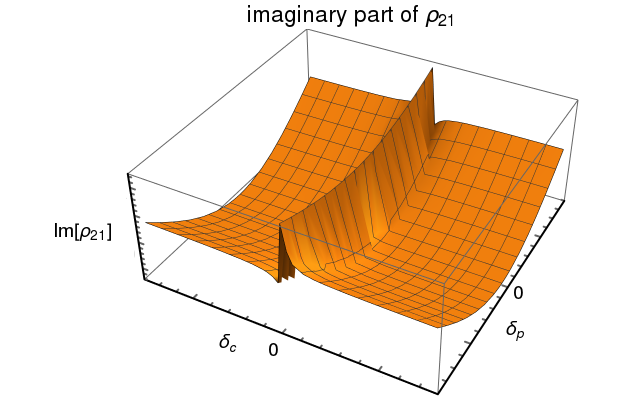}
	\caption{Schematic image of the line profile corresponding to $Im\{\rho_{21}\}$, Eq. (\ref{4}), for different values of detunings $\delta_p$ and $\delta_c$.}
	\label{Fig3}
\end{figure}

However, the series expansion in Rabi frequencies cannot be employed in our case due to the smallness of level width $\Gamma_2\approx 2.85\cdot 10^{-15}$ $s^{-1}$. Nonetheless, the imaginary part of $\rho_{21}$ can be separated out
\begin{eqnarray}
\label{4}
-Im\{\rho_{21}\}\equiv -Im\{\rho_{21}^{(1)}\}-Im\{\rho_{21}^{(2)}\}=
\\
\nonumber
\frac{\gamma_{21}\Omega_p/2 }{\left(\delta_p-\frac{(\delta_p+\delta_c)\Omega_c^2/4}{(\delta_p+\delta_c)^2+\gamma_{31}^2}\right)^2+\left(\gamma_{21}+\frac{\gamma_{31}\Omega_c^2/4}{(\delta_p+\delta_c)^2+\gamma_{31}^2}\right)^2}
+
\\
\nonumber
\frac{\gamma_{31}\Omega_p \Omega_c^2/8}{\left[\left(\delta_p-\frac{(\delta_p+\delta_c)\Omega_c^2/4}{(\delta_p+\delta_c)^2+\gamma_{31}^2}\right)^2+\left(\gamma_{21}+\frac{\gamma_{31}\Omega_c^2/4}{(\delta_p+\delta_c)^2+\gamma_{31}^2}\right)^2\right]((\delta_p+\delta_c)^2+\gamma_{31}^2)}.
\end{eqnarray}
The first term here represents the one-photon $|1\rangle\rightarrow|2\rangle$ (21 cm) absorption process, the second term can be associated with the additional process $|1\rangle\rightarrow|2\rangle\rightarrow|3\rangle\rightarrow|2\rangle$ \cite{Wiel}. In absence of the second field $\Omega_c=0$, the second term in Eq. (\ref{4}) vanishes, and the ordinary definition (\ref{3a}) can be found. The line profiles corresponding to Eqs. (\ref{3a}) and (\ref{4}) are given schematically in Figs.~\ref{Fig2} and \ref{Fig3}, respectively. In particular, Figs.~\ref{Fig2} and \ref{Fig3} show that the contribution arising via the additional pathways leads to the distortion the line profile in the vicinity of zero detuning $\delta_c$.

%In case of $\Omega_c\rightarrow 0$, the second term in Eq. (\ref{4}) can be considered as the correction to the first one. However, in most events, the influence of the additional transitions is significant, and expression (\ref{4}) represents the two competitive terms. Smallness of levelwidth $\gamma_{21}=\Gamma_2$ requires the accurate analysis of the expression $Im\{\rho_{21}\}$ also. 
Thus, the absorption coefficient and the optical depth, respectively, cannot be described by the single Lorentz contour Eq. (\ref{3a}) with the subsequent transformation to the Voigt profile. The Voigt fitting, in this case, is the overabundant and covers the physical processes occurring in the medium illuminated with the radiation from a powerful light source at the adjacent resonances. It can be noted also that the first term corresponding to the absorption at 21 cm line ($|1\rangle\rightarrow|2\rangle$ transition) shows the line profile to be broadened and shifted {\it a priori}.

The dimensionless correction to the optical depth Eq. (\ref{3}) arising in context of Eq. (\ref{4}) can be defined as follows
\begin{eqnarray}
\label{5}
\tau=\tau_0(1+\delta\tau)\, ,
\end{eqnarray}
where $\tau_0$ corresponds to single profile $Im\{\rho_{21}^{(1)}\}$ %or $Im\{\rho_{21}^{(2)}\}$ depending on the contribution of $Im\{\rho_{21}^{(1)}\}$ and $Im\{\rho_{21}^{(2)}\}$, respectively.
and correction $\delta\tau$ is
\begin{eqnarray}
\label{6}
\delta\tau = \frac{Im\{\rho_{21}\}-Im\{\rho_{21}^{(1)}\}}{Im\{\rho_{21}^{(1)}\}}\equiv\frac{Im\{\rho_{21}^{(2)}\}}{Im\{\rho_{21}^{(1)}\}}.
%\\
%\nonumber
%\delta\tau_2 = \frac{Im\{\rho_{21}\}-Im\{\rho_{21}^{(2)}\}}{Im\{\rho_{21}^{(2)}\}}\equiv\frac{Im\{\rho_{21}^{(1)}\}}{Im\{\rho_{21}^{(2)}\}}.
\end{eqnarray}
Expression (\ref{6}) simplifies to
\begin{eqnarray}
\label{7}
\delta\tau=\frac{\gamma_{31}}{\gamma_{21}}\frac{\Omega_c^2}{4((\delta_p+\delta_c)^2+\gamma_{31}^2)}.
%\xlongequal{\delta_{p}+\delta_c= 0}\frac{\Omega_c^2}{4\gamma_{21}\gamma_{31}}
%\,\,\underset{\delta_{p}+\delta_c\rightarrow 0}{ =\joinrel= \xlongequal{}}aaa                        \xlongequal{\text{def}}
\end{eqnarray}
%which is equal to $\delta\tau_2=\frac{\Omega_c^2}{4\gamma_{21}\gamma_{31}}$ at zero detunings.
Note that the expression (\ref{7}) is of a resonant nature but independent of probe field $\Omega_p$ acting on the $21$ cm line.

\section{None-Doppler broadening and frequency shift}

The section deals with the effects of absorption line broadening and frequency shift for an atom at rest and assumption of $\Omega_c$ smallness.
\subsection{None-Doppler broadening}

The non-Doppler line broadening for the $|1\rangle \rightarrow |2\rangle$ transition follows from the denominator in the first term and is proportional to $\Omega_c$. This broadening can be expressed as an additional term to natural width $\gamma_{21}$:
\begin{eqnarray}
\label{8}
\gamma=\gamma_{21}+\frac{\gamma_{31}\Omega_c^2/4}{(\delta_p+\delta_c)^2+\gamma_{31}^2}=\gamma_{21}+\gamma_{{\rm broad}}.
\end{eqnarray}
The maximum value of $\gamma_{{\rm broad}}$ is attained for the exact two-photon resonance $\delta_p+\delta_c=0$:
\begin{eqnarray}
\label{9}
\gamma_{{\rm broad}}=\frac{\Omega_c^2}{4\gamma_{31}}.
\end{eqnarray}
For very powerful light source and small distances between the absorber and the source, it can be expected that the value of $\gamma_{{\rm broad}}$ is possibly larger than natural level width $\gamma_{21}$.%Value of $\gamma_{{\rm broad}}$ is smaller than $\gamma_{21}$ in case of $\Omega_c^2\leq \gamma_{31}\gamma_{21}$. Such situation can be achieved for absorbers, which are very distant from the source of field or the radiation intensity for $|2\rangle \rightarrow |3\rangle$ line is suppressed with respect to the $|1\rangle \rightarrow |2\rangle$ transition, see Eqs. (\ref{1}). In other cases, $\gamma_{{\rm broad}}$ can contribute substantially.

Taking into account the motion of interstellar gas cloud, we can find that the resonant frequency should be shifted. This Doppler shift leads to $\delta_c\rightarrow\delta_c+\frac{\upsilon}{c}\omega_c$ \cite{Wiel}, where $c$ is the speed of light. The speed of hydrogen clouds can be on the order of few hundreds $km\cdot s^{-1}$ \cite{Fox}, \cite{Moller} and, in some cases, as high as thousand kilometers per second \cite{Gupta}. Then the sum of detunings $\delta_p+\delta_c$ can be estimated as $(10^{-3}-10^{-2})\omega_c\sim (10^4-10^5)\gamma_{31}$, where $\gamma_{31} = \frac{1}{2}\Gamma_{2p} = \frac{1}{2}6.265\cdot 10^8$ $s^{-1}$ and Ly$_\alpha$ frequency is $\omega_c = 2.466\cdot 10^{15}$ $s^{-1}$. Thus, the Doppler shift leads to the suppression of $\gamma_{{\rm broad}}$. Nonetheless, since the emission spectrum of the source is of a continuum nature, the case of the exact two-photon resonance can always be singled out. It should be underscored that this discussion corresponds to $\gamma_{{\rm broad}}$ and does not cancel the Doppler broadening leading to the Voigt profile.

\subsection{Frequency shift}

Equation (\ref{4}) allows also finding the frequency shift for the transition $|1\rangle \rightarrow |2\rangle$. To this end, detuning $\delta_p$ can be considered as the 'scanning' parameter (variable). Then the resonance condition reads
\begin{eqnarray}
\label{10}
\delta_p-\frac{(\delta_p+\delta_c)\Omega_c^2/4}{(\delta_p+\delta_c)^2+\gamma_{31}^2}=0.
\end{eqnarray}
Now, the frequency shift is zero for the exact two-photon resonance, $\delta_p+\delta_c=0$. In case, when the detuning of the two-photon resonance $\delta_p+\delta_c=\gamma_{31}/2$, the frequency shift can be found as 
\begin{eqnarray}
\label{10a}
\delta_{{\rm shift}}=\frac{\Omega_c^2}{4\gamma_{31}}.
\end{eqnarray}
Here, level width $\gamma_{31}$ acts as a natural parameter for the atomic resonant excitation.

Another result arises in the assumption that $\delta_p\sim\gamma_{21}\ll \delta_c\sim\gamma_{31}$ (one-photon resonances). Then, neglecting $\delta_p$ in the second term of Eq. (\ref{10}), the frequency shift is
\begin{eqnarray}
\label{11}
\delta_{{\rm shift}}=\frac{\delta_c\Omega_c^2}{4\delta_c^2+4\gamma_{31}^2}.
\end{eqnarray}
Here, we can take into account the motion of hydrogen cloud by using parameter $\beta$: $\delta_c=\upsilon/c\cdot\omega_c\equiv\beta\gamma_{31}$. Therefore, 
\begin{eqnarray}
\label{12}
\delta^{\beta}_{{\rm shift}}=\frac{\Omega_c^2}{4\gamma_{31}}\frac{\beta}{1+\beta^2}\approx \frac{\Omega_c^2}{4\gamma_{31}}\frac{1}{\beta}.
\end{eqnarray}
The shift is negligibly small, and the maximum shift can be attained for the two-photon resonance with the detuning being $\delta_p+\delta_c=\gamma_{31}/2$, see Eq. (\ref{10a}).
%\begin{eqnarray}
%\label{13}
%\frac{d}{d\delta_p}\left(Im\{\rho_{21}\}\right)=0\, ,
%\\
%\nonumber
%\frac{d}{d\delta_c}\left(Im\{\rho_{21}\}\right)=0.
%\end{eqnarray}
%Further on, the numerical evaluation of the frequency shift, broadening and corrections to the optical depth, which arise in ISM, are discussed.

\section{Numerical results}
\label{nr}

To evaluate the contribution of non-Doppler broadening, frequency shift and correction to the optical depth, see Eqs. (\ref{9}), (\ref{11}) and (\ref{7}), respectively, one is to find Rabi frequency $\Omega_c$. It can be done via the flux density or luminosity of the light sources and the distance between the source and the absorber. To this end, the observation data of Damped Lyman-$\alpha$ systems at $1216$ \AA\, line in hydrogen \cite{Gupta}, \cite{Cur}-\cite{Wolf} were used. Finding the distance employed the following expression:
\begin{eqnarray}
\label{14}
r=\frac{c}{H_0}(z_{{\rm em}}-z_{{\rm abs}})\, ,
\end{eqnarray}
where $H_0=2.3\cdot 10^{-18} s^{-1}$ is the Hubble constant, $z_{{\rm em}}$, $z_{{\rm abs}}$ are the redshifts of the source and the absorber, respectively. The radiation intensity at the absorber can be defined as
\begin{eqnarray}
\label{15}
I_{{\rm abs}}=\frac{L_*}{4\pi(1+z_{{\rm em}}-z_{{\rm abs}})^4r^2}\, ,
\end{eqnarray}
where $L_*$ is the star luminosity (measured in units of $W/Hz$), which is independent of the distance. For observed flux density $S$ at frequency $\nu$, the intensity at the absorber is
\begin{eqnarray}
\label{16}
S_{{\rm abs}}=S_0\,\nu_0\frac{z_{{\rm em}}^2 (1 + z_{{\rm em}})^4}{(z_{{\rm em}}-z_{{\rm abs}})^2 (1 + z_{{\rm em}}-z_{{\rm abs}})^4}\, ,
\end{eqnarray}
where $S_0$ is the measured flux density and $\nu_0$ is the frequency of the corresponding transition. %Although the bandwidth can be in the range of $2-16 MHz$ \cite{Cody}, \cite{Cur2012} our numerical estimations used the value $5 MHz$. The value does not correspond to all the cases but can serve for the evaluation of the order of the arising corrections with good accuracy.

The flux density for the Ly$_{\alpha}$ line can be expressed via electric field strength $E_{\alpha}$ as
\begin{eqnarray}
\label{17}
S_{\alpha}=\frac{1}{2}\sqrt{\frac{\epsilon_0}{\mu_0}}c E_c^2\, ,
\end{eqnarray}
where $\epsilon_0$ is the vacuum permittivity and $\mu_0$ is the vacuum permeability. In principle, the Rabi frequency for the $21$ cm transition can be defined in the same way, i.e. as $S_{21}=\frac{1}{2}\sqrt{\frac{\epsilon_0}{\mu_0}}c B_p^2$. The data used in our calculations are collected in Table 1.

\begin{table}
\tiny{
\caption{ The first column contains the names of sources. The second and third columns represent the redshift of the star and the absorber, respectively, where the $21$ cm absorption in conjunction with Ly$_\alpha$ absorption were observed. The next column shows the density flux at $1.4$ $GHz$ frequency (hyperfine splitting of the ground state in hydrogen atom, $|1\rangle\leftrightarrow|2\rangle$ transition in the present calculations). The fifth column lists the values of density flux and luminosity at Ly$_\alpha$ frequency. The values of the hydrogen velocity at $21$ cm line are given in the sixth column. %The last but one column represents the values of the corresponding column densities of hydrogen atoms.
 Finally, the optical depth values are given in the last column of the table. References with the data used are given in square brackets.
\\ 
${}^a$ The component with the poorest accuracy is taken from \cite{Ish}. 
\\
${}^b$ Just one of components from \cite{Lab} is considered.
\\
${}^c$ Cloud 3 is treated in accordance to data \cite{Wolf}.
\\
${}^d$ optical depth at the Lyman-limit \cite{Led}.}
\begin{center}
\begin{tabular}{c|c|c|c|c|c|c}
\hline
\hline
Name & $z_{{\rm em}}$ & $z_{{\rm abs}}$ & $S_{1.4GHz}$, $Jy$ & $logL_{\alpha}$, $(W\cdot Hz^{-1})$ & $\upsilon$, $(km\cdot s^{-1})$ %& $N_{HI}$, $cm^{-2}$
 & $\tau_0$  \\

\hline
\hline

0235+164 & 0.94 & 0.523869 & $1.7 \cite{Cur}$ & $log[\nu f]\approx -12.5 \cite{Abdo}$ $\frac{erg}{cm^2 s}$ & $125 \cite{Kane}$ %& $(4.5\pm 0.4)\cdot 10^{21} \cite{Tur}$ 
& $(\frac{1}{f})\int\tau d\upsilon=13\pm 0.6 \cite{Kane}$  \\

  \hline
3C 190 & 1.1946 & 1.19565 & $2.47 \cite{inet}$ & $0.17 Jy \cite{inet}$ & $-37.1 \cite{Ish}$ %& $\approx 2\cdot 10^{20} \cite{inet}$ 
& $0.0027\pm 0.0002 \cite{Ish}^a$  \\

  \hline
3C 216 & 0.668 & 0.63 & $3.4 \cite{inet}$ & $22.699 \cite{Curran2008}$ & $102  \cite{Ver,Gupta}$ %& $1.30\cdot 10^{20} \cite{Gupta}$ 
& $0.38 \cite{Pih}$  \\

  \hline
J0414+0534 & 2.6365 & 0.9586 & $1.82 \cite{Tanna}$ &  $22.188 \cite{Curran2008}$ & $205 \cite{Curran2007}$  %& $1.60(4)\cdot 10^{18}(\frac{T_s}{f}) \cite{Curran2007}$ 
& $0.0212(16) \cite{Curran2007}$  \\

  \hline
J0414+0534 & 2.6365 & 2.63534 & $3.31 \cite{Moor}$ &  $22.188 \cite{Moor}$ & $-94 \cite{Moor}$  %& $(4.5\pm 1.0)\cdot 10^{18}(\frac{T_s}{f}) \cite{Moor}$ 
& $(0.015\pm 0.002) \cite{Moor}$  \\

  \hline
0902+343 & 3.398 & 3.3968 & $1.2 \cite{Cody}$ & $22.422 \cite{Curran2008}$ & $120 \cite{Cody,Ch}$ %& $3\cdot 10^{21} \cite{Cody}$ 
&  --  \\

  \hline
3C 49 & 0.621 & 0.6207 & $7.28 \cite{Lab}$ & $20.777 \cite{Curran2008}$ & $-138 \cite{Lab}$ %& $(1.5\pm 0.3)\cdot 10^{20}(\frac{T_s}{100K}) \cite{Lab}$ 
& $0.036\pm 0.003 \cite{Lab}^b$  \\

  \hline
3C 286 & 0.849 & 0.692153 & $14.7 \cite{inet}$ & $2.7 \, Jy\, \cite{inet}$ & $4.2 \cite{Wolf}$ %& $log[N_{HI}]=21.25\pm 0.02 \cite{Boisse}$ 
& $0.280\pm 0.004 \cite{Wolf}^c$  \\

  \hline
0118-272 & 0.559 & 0.558  & $0.93 \cite{inet}$ & $0.95\, Jy\, \cite{inet}$ & -- & $log[N_{HI}]=20.3 \cite{inet}$  \\

  \hline
0405-331 & 2.570 & 2.562  & $0.63 \cite{MM}$ & $0.56\, Jy\, \cite{MM}$ & -- & $log[N_{HI}]=20.6 \cite{MM}$  \\

  \hline
0537-286 & 3.104 & 2.976  & $0.862 \cite{inet}$ & $0.90\, Jy\, \cite{inet}$ & -- & $(0.41\pm0.22)\cdot 10^{22}$  \\

  \hline
0957+561A & 1.413 & 1.391  & $0.59 \cite{inet}$ & $0.15\, Jy\, \cite{inet}$ & 25 \cite{Mich} & $N_{HI}=7\cdot 10^{19}\pm 30\%$ \cite{Turn} \\

  \hline
0248+430 & 1.31 & 0.3939  & $1.4 \cite{inet}$ & $1.5\, Jy\, \cite{inet}$ & 40 \cite{Rao} & $N_{HI}=(3.6\pm 0.4)\cdot 10^{19}$ \\

  \hline
0336-017 & 3.197 & 3.0619  & $0.60 \cite{inet}$ & $0.15\, Jy\, \cite{inet}$ & 13  & $log[N_{HI}]=21.25$, $\tau_0< 0.2$  \cite{Tara}\\

  \hline
0528-250 & 2.813 & 2.8110  & $1.16 \cite{inet}$ & $0.59\, Jy\, \cite{inet}$ & $5$ \cite{Sri} & $log[N_{HI}]=21.3\pm 0.1$ \cite{Lu}\\

  \hline
2128-123 & 0.501 & 0.430  & $1.8 \cite{inet}$ & $0.7\, Jy\, \cite{inet}$ & $75$ \cite{Led} & $log[N_{HI}]=19.37\pm 0.08$, $\tau_{LL}\simeq 150 \cite{Led}^d$ \\
%  \hline
%2059-360 & 3.090 & 3.0825  & $0.003 \cite{inet}$ & $\approx 0.28\, Jy \cite{Condon}$ & -- & $log[N_{HI}]=20.98\pm 0.08 \cite{Condon}$   \\

  \hline
  \hline
\end{tabular}
\end{center}
%The velocity width $\Delta V$ quoted here is the entire velocity range over which absorption is seen \cite{Kane}
% \\
%The peak of the absorption occurs at a velocity of $-210 km\cdot s^{-1}$\cite{Ish}
%!!!!!!!2-MHz bandwidth Curran 2008!!!!!!!!!!!!
%
%Bandwidth 5 MHz \cite{Cody}
%\\
%The velosity offset of the detected line ... \cite{Ver}
%\\
%The resulting HI-optical rest frame velocity offset $\upsilon_{HI}-\upsilon_{opt}$ \cite{Curran2007}
%\\
%$\upsilon$ is the velocity offset relative to redshift 2.63647
%\\
%rest frame velocity $120 km \cdot s^{-1}$ \cite{Cody} ---
%\\
%central wavelength \cite{Lab}
%\\
%several components spread over $160 km s^{-1}$ \cite{Boisse}
}
\end{table}

Employing equations (\ref{15})-(\ref{17}) and data from Table 1, the line broadening and the frequency shift were evaluated with Eq. (\ref{9}) and Eq. (\ref{12}), respectively. The correction to optical depth $\delta\tau$ for the zero and non-zero detunings are calculated with Eq. (\ref{7}). The Doppler effect can be taken into account with the use of data in Table 1 (sixth column). It should be pointed out that in view of smallness of frequency $\omega_{21}$ in respect to Ly$_\alpha$, detuning $\delta_p$ can be set equal to zero, since $\delta_p=\beta\omega_p\ll\delta_c=\beta\omega_c$. The results of the numerical calculations are given in Table 2, notations $\delta\tau_0$ and $\delta\tau$ corresponding to the zero and non-zero detunings, respectively.

\begin{table}
\caption{The first column contains the name of sources in correspondence to the Table 1. The second column gives the values evaluated with Eq. (\ref{9}) for the non-Doppler broadening. Frequency shift $\delta_{\rm shift}$ Eq. (\ref{12}) for the $|1\rangle \leftrightarrow|2\rangle$ transition is represented in the third column. The fourth column lists the relative contributions of $\delta \tau$ at detunings $\delta_p\ll\delta_c=\frac{\upsilon}{c}\omega_{32}$. Values of $\delta\tau_0$ are given in the last column, in which the second subrow contains the values of correction to the optical depth listed in Table 1.}
	\begin{center}
		\begin{tabular}{c|c|c|c|c}
			\hline
			\hline
			Name & $\gamma_{\rm broad}$ in $s^{-1}$ Eq. (\ref{9}) & $\delta^\beta_{\rm shift}$ in $s^{-1}$  &  $\delta\tau$ &  $\delta\tau_0$ \\
			& $vs$ $\Gamma_2=2.85\cdot 10^{-15}$ $s^{-1}$ & Eq. (\ref{12}) & at $\delta_c=\frac{\upsilon}{c}\omega_{32}$ & $\delta\tau_0\cdot \tau_0$ \\
			\hline
			\hline
			
 0235+164 & $9.28\cdot 10^{-20}$ & $3.87\cdot 10^{-23}$ & $1.53\cdot 10^{-13}$ & $6.51\cdot 10^{-5}$\\
 &  &  &  & $8.46\cdot 10^{-4}$\\

			\hline
 3C 216 & $2.34\cdot 10^{-17}$ & $7.97\cdot 10^{-21}$ & $5.805\cdot 10^{-11}$ & $0.0164$\\
  &  &  &  & $6.23\cdot 10^{-3}$\\
			\hline
			
 J0414+0534 & $4.93\cdot 10^{-24}$ & $3.37\cdot 10^{-27}$ &  $3.02\cdot 10^{-18}$ & $3.46\cdot 10^{-9}$ \\
			$z_{abs}=0.9586$ & &  &  & $7.33\cdot 10^{-11}$\\
			\hline
 J0414+0534 & $5.59\cdot 10^{-16}$ & $1.75\cdot 10^{-19}$ & $1.63\cdot 10^{-9}$ & $0.392$ \\
			$z_{abs}=2.63534$ & & & & $5.88\cdot 10^{-3}$\\
			\hline

 0902+343 & $8.502\cdot 10^{-16}$ & $3.403\cdot 10^{-19}$ & $1.52\cdot 10^{-9}$ & $0.597$ \\
			\hline
			
 3C 49 & $3.07\cdot 10^{-16}$ & $1.41\cdot 10^{-19}$ &  $4.15\cdot 10^{-10}$ & $0.215$\\
  &  &  &  & $7.74\cdot 10^{-3}$\\ 
          
 			\hline
 0248+430 & $6.38\cdot 10^{-16}$ & $8.51\cdot 10^{-20}$ &  $1.03\cdot 10^{-8}$ & $0.448$\\
    &  &  &  & --\\
              
 			\hline
 2128-123 & $1.32\cdot 10^{-14}$ & $3.31\cdot 10^{-18}$ & $6.07\cdot 10^{-8}$ & $0.108$\\
    &  &  &  & $16.1$\\
  			\hline\hline\Xhline{2\arrayrulewidth}
			
 3C 190 & $3.08\cdot 10^{-11}$ & $3.798\cdot 10^{-15}$ & $5.79\cdot 10^{-4}$ & $4.63\cdot 10^{-5}$ \\
  &  &  &  & $1.25\cdot 10^{-7}$\\
  			\hline
			
 3C 286 & $8.398\cdot 10^{-14}$ & $1.18\cdot 10^{-18}$ &  $1.23\cdot 10^{-4}$ & $0.0169$\\
  &  &  &  & $0.00475$\\
  
			\hline
 0118-272 & $1.72\cdot 10^{-10}$ & -- & -- & $8.29\cdot 10^{-6}$ \\
   &  &  &  & --\\
 
 			\hline
 0405-331 & $8.95\cdot 10^{-10}$ & -- &  -- & $1.59\cdot 10^{-6}$\\
    &  &  &  & --\\
  
 			\hline
 0537-286 & $2.07\cdot 10^{-11}$ & -- &  -- & $6.88\cdot 10^{-5}$\\
    &  &  &  & --\\
      
 			\hline
 0957+561A & $1.89\cdot 10^{-12}$ & $1.58\cdot 10^{-16}$ &  $7.81\cdot 10^{-5}$ & $7.53\cdot 10^{-4}$\\
    &  &  &  & $0.0753$\\
      
 			\hline
 0336-017 & $3.69\cdot 10^{-12}$ & $1.60\cdot 10^{-16}$ &  $5.63\cdot 10^{-4}$ & $3.86\cdot 10^{-4}$\\
    &  &  &  & $7.72\cdot 10^{-5}$\\
      
 			\hline
 0528-250 & $2.41\cdot 10^{-8}$ & $4.02\cdot 10^{-13}$ & $0.0402$ & $5.91\cdot 10^{-8}$\\
    &  &  &  & --\\

%			\hline
% 2059-360 & $1.53\cdot 10^{-9}$ & -- &  -- & $9.32\cdot 10^{-7}$\\	
%    &  &  &  & --\\	

			\hline
			\hline
		\end{tabular}
	\end{center}
\end{table}

\section{Analysis of results}

Typically, astrophysical investigations of absorption lines employ the one-photon profile. Within the framework of density matrix formalism, the one-photon absorption line profile can be derived in the two-level approximation, see Eq. (\ref{3a}), for the atom in an external field. According to the approximation, the monochromatic absorption coefficient and, hence, the optical depth can be defined via the imaginary part of the density matrix element, Eqs. (\ref{2}), (\ref{3}). In this case, the definition of the optical depth and all the ensuing physical quantities leads to the results of the 'ordinary' theory. However, section~\ref{opt} demonstrated that this was the case of the zeroth approximation, when the one-photon absorption process is considered for an isolated transition in the atom. Accounting for the absorption/emission processes that occur at adjancent transitions leads to a substantial modification of the line profile. The paper employs the three-level approximation for the atom. Within the theory \cite{SDP}-\cite{Wiel}, the imaginary part of a density matrix element fails to correspond to the isolated transition and strongly depends on field parameters defined for the adjacent resonance, Eq. (\ref{1a}).

\subsection{Correction to the optical depth}

The absorption line profile of the $|1\rangle\leftrightarrow |2\rangle$ transition is shown in the paper to be formed by two contributions: $Im\{\rho_{21}^{(1)}\}$ and $Im\{\rho_{21}^{(2)}\}$, see Eq. (\ref{4}). The additional term in the line profile is proportional to $\Omega_p$ and $\Omega_c$, Rabi frequencies of the $|1\rangle\leftrightarrow |2\rangle$ and $|2\rangle\leftrightarrow |3\rangle$ transitions, respectively. Its physical interpretation was given in \cite{Wiel} as the interfering emission/absorption pathways in an atom. Using such modification, the correction to the optical depth can be found as Eq. (\ref{7}). The correction should be small for $\Omega_c\ll\gamma_{21}$. However, in view of smallness of level width $\gamma_{21}$, the condition is met for very distant source and absorber, while the opposite situation can be found for $z_{\rm em}\approx z_{\rm abs}$ and a very powerful source of light. In this case, the main contribution to the line profile comes from the second term $Im\{\rho_{21}^{(2)}\}$, and the correction to the optical depth should be taken as the reciprocal value of the former one. Numerical results for the correction to the optical depth at zero detunings in case of $\Omega_c\ll\gamma_{21}$ and $\gamma_{21}\ll\Omega_c$ are collected in the first and second parts of Table 2 as $\delta\tau_0$, respectively.

Although the case of zero detuning can be always singled out, since the light source emission is of the continuum nature, the velocity of clouds can be taken into account for the detailed description of the 21 cm absorption line profile ($|1\rangle\leftrightarrow |2\rangle$ transition) in the interstellar medium. This can be rendered by the approximate equality $\delta_p+\delta_c\approx \frac{\upsilon}{c}\omega_{32}$, where values of $\upsilon$ are listed in Table 1. Numerical results for $\delta\tau$ are also given in Table 2. 

In particular, it follows from Table 2 that the contribution of $\delta\tau_0$ can be significant and exceed the accuracy of the experimental determination of $\tau_0$. %In all the other cases, the correction Eq. (\ref{7}) never exceeds the level of $0.04\%$. However, we found two events where the contribution of the first term was larger: J0414+0534 and $z_{abs}=0.9586$ and 3C 286. For 3C 286 and the non-zero detunings, the second term $Im\{\rho_{21}^{(2)}\}$ contributes at the level of $10.5\%$ with respect to $Im\{\rho_{21}^{(1)}\}$ and is negligible for J0414+0534 and $z_{abs}=0.9586$. The most important conclusion ensues in the context of the hydrogen column density determination. Values of the optical depth and column density $N_{HI}$ are overestimated. 
Although our analysis is rather rough and does not include the Voigt profile fitting, the main conclusion is that the two-level approximation of atom is insufficient. Already in the three-level approximation, the additional processes occurring in the atom should be taken into consideration in the appropriate fitting of the absorption profile. The parameters of medium extracted from such fitting can be corrected with use of Eqs. (\ref{4}), (\ref{7}).

\subsection{None-Doppler broadening and frequency shift}

In keeping with Eq. (\ref{4}), the absorption line profile can be analyzed in terms of line broadening. Without regard to which contribution is dominant, $Im\{\rho_{21}^{(1)}\}$ or $Im\{\rho_{21}^{(2)}\}$, the absorption line derived via the density matrix element is modified by width $\gamma_{\rm broad}$, Eq. (\ref{8}). The maximum broadening can be estimated as $\Omega_c^2/4\gamma_{31}$, see Eq. (\ref{9}). The values of $\gamma_{\rm broad}$ at zero detunings are given in Table 2. It is found that the broadening can be significant and exceed the natural line width by several orders of magnitude. %With the account for motion of hydrogen cloud, smaller values of $\gamma_{\rm broad}$ can be obtained. These values can be estimated by the similar expression as for the frequency shift.

An aspect of interest in such investigations consists in determination of frequency shift and, therefore, refinement of the distances to the source of light and the sizes of cloud. The accuracy of the redshift determination is on the level of $10^{-10}$ \cite{Darl}, and reaches $10^{-11}$ in some cases \cite{Wolf}. The procedure of the redshift definition can be reduced to finding the maximum of the corresponding line contour. In the same way, the frequency shift was obtained, Eq. (\ref{12}). The maximum frequency shift arises when $\delta_p+\delta_c=\gamma_{31}/2$. Numerical values are listed in Table 2 in the third column for the $\delta^\beta_{\rm shift}$ and are $\gamma_{\rm broad}$ at maximums (see the second column of Table 2). So, the uncertainty of the redshift, $\delta z_{\rm shift}$, can be estimated via frequency shift $\delta_{\rm shift}$:
%\begin{eqnarray}
%\label{18}
%\nu_{\rm obs}=\frac{\nu_0+\delta_{\rm shift}}{1+z_{\rm abs}+\delta z_{\rm shift}}\, ,
%\end{eqnarray}
%where $\nu_{\rm obs}$ is the observed frequency, $\nu_0$ is the resonant frequency of the corresponding transition and $\delta z_{\rm shift}$ represents the redshift uncertainty. The equation can be rewritten as
%\begin{eqnarray}
%\label{18a}
%\nu_{\rm obs}(1+z_{\rm abs}) = \frac{\nu_0+\delta_{\rm shift}}{1+\frac{\delta z_{\rm shift}}{1+z_{\rm abs}}}\approx \nu_0-\nu_0\frac{\delta z_{\rm shift}}{1+z_{\rm abs}}+\delta_{\rm shift}-\delta_{\rm shift}\frac{\delta z_{\rm shift}}{1+z_{\rm abs}}.
%\end{eqnarray}
%The last term here is negligible, and $\nu_{\rm obs}(1+z_{\rm abs})\approx\nu_0$. Then
\begin{eqnarray}
\label{19}
\delta z_{\rm shift} = \frac{\delta_{\rm shift}(1+z_{\rm abs})}{\nu_0},
\end{eqnarray}
where $\nu_0$ is the transition frequency. The values given in Table 2 show that this effect is quite negligible and can be excluded from the corresponding analysis.

\section{Conclusions}

The paper studied the 21 cm line profile for the hydrogen atom within the framework of the density matrix formalism. Application of density matrix theory allows the detailed description of the emission/absorption processes when the atom is illuminated by the powerful source of light. The one-photon absorption line profile can be obtained in this case within the two-level approximation of atomic system, which represents the zeroth approximation. However, the additional emission/absorption processes should be taken into account. These processes can be evaluated within the three-level approximation. The additional interfering transitions were shown to lead to a substantial modification of the corresponding line profile. % Moreover, the Doppler effect can be partially taken into account by substitution $\delta_p\rightarrow \frac{\upsilon}{c}\omega_{21}$ and $\delta_c\rightarrow \frac{\upsilon}{c}\omega_{32}$.
%The 21 cm absorption line profile in hydrogen atom was analysed with the account for adjacent Ly$\alpha$ resonance. Within the three-level approximation 

Corrections to the frequency and level width were found. The frequency shift can be attributed to the redshift. Although the frequency shift is negligibly small, the width of line profile can be several orders larger than the natural one. The most significant effect arises for the optical depth. In particular, uncertainty of the optical depth determination is about $13\%$ in case of J0414+0534 source, see Table 1, whereas correction Eq. (\ref{7}) is on the order of $39\%$. The same result can be found for the 3C 49 light source: uncertainty and correction are about $8\%$ and $20\%$, respectively. The magnitude of correction to the optical depth and, therefore, to the column density can be as high as $60\%$, see Table 2. In particular, when $z_{\rm}\approx z_{\rm abs}$, fitting of the observed line profile with the one-photon isolated resonant contour can lead to the overestimation of the corresponding magnitudes.

\ack{This work was supported by Russian Science Foundation (grant 17-12-01035).}

\end{document}